\documentclass[prd,aps,floats,twocolumn,nofootinbib,superscriptaddress]{revtex4-2}
\usepackage{natbib}
\usepackage{commath}
\pdfoutput=1
\usepackage[usenames, dvipsnames]{color}
\usepackage[T1]{fontenc}
\usepackage{lmodern}
\usepackage{graphicx, array}
\usepackage{multirow}
\usepackage{graphics}
\usepackage{graphics,epsfig}
\usepackage{amssymb}
\usepackage{amsmath}
\usepackage{ulem}
\usepackage{multirow}
\usepackage{subfigure}
\usepackage{bm}
\usepackage{txfonts}
\usepackage{cancel}
\usepackage{url}
\usepackage{hyperref}
\usepackage{lipsum}
\usepackage{graphicx}
\usepackage{comment} 
\usepackage{soul}

\newcolumntype{C}[1]{>{\centering\arraybackslash}m{#1}}

\def\be{\begin{equation}}
\def\ee{\end{equation}}
\def\bi{\begin{itemize}}
\def\ei{\end{itemize}}
\def\ben{\begin{enumerate}}
\def\een{\end{enumerate}}
\def\bt{\begin{tabular}}
\def\et{\end{tabular}}
\def\bc{\begin{center}}
\def\ec{\end{center}}

\def\bea{\begin{eqnarray}}
\def\eea{\end{eqnarray}}

\def\ba{\begin{eqnarray}}
\def\ea{\end{eqnarray}}

\renewcommand{\arraystretch}{1.2}

\let\oldhat\hat

\renewcommand{\hat}[1]{\oldhat{\boldsymbol{\mathbf{#1}}}}

\graphicspath{ {./images/},{ ./images/ratioFigs/},{ ./images/windowFigs/},{ ./Version1/}}

\begin{document}

\title{Full Nonlinear Velocity Reconstruction With Transformer and Ensemble Tree Machine Learning}

\author{Yulin Gong}
\affiliation{Department of Astronomy, Cornell University, Ithaca, NY 14853, USA}

\author{Rachel Bean}
\affiliation{Department of Astronomy, Cornell University, Ithaca, NY 14853, USA}

\date{\today}

\begin{abstract}
Accurate reconstruction of peculiar velocities from galaxy positions is important for probing the motion and evolution of large scale structure. They are sensitive to the cosmological effects of gravity and dark sector matter, and complement other velocity inference methods such as the kinematic Sunyaev-Zel'dovich (kSZ) imprinted in the CMB. 
We show that machine learning methods improve velocity reconstruction by capturing nonlinear contributions. Specifically, we train both a gradient boosting decision tree (GBDT) and a Transformer using multi-scale features to predict the residual between the actual velocity and the estimate from linear theory for both the line-of-sight and transverse components. 
We evaluate our approach in both periodic box and, more realistic, lightcone settings using mock galaxy catalogs from the \textsc{AbacusSummit} simulations tailored to DESI spectroscopic surveys of luminous red galaxies (LRGs) and emission line galaxies (ELGs). We also assess the impact of redshift uncertainties such as those in Rubin LSST photometry. 
Both models significantly outperform the linear theory with the Transformer achieving the best performance. They more accurately recover the velocity power spectrum and maintain a higher cross-correlation with the true velocities across a wider range of spatial scales.  
Finally, we demonstrate two applications relevant to kSZ analyses: estimating cluster pairwise velocity correlations and stacked cluster density profiles. This machine learning framework for nonlinear velocity reconstruction  opens up powerful new applications of survey data from DESI, Rubin LSST, Euclid and the Roman Space Telescope.
\end{abstract}

\maketitle
 
\section{Introduction}
\label{sec:intro}
Just as the perturbations of planetary motions provide precise tests of General Relativity and help map the distribution of mass in the Solar System, so the peculiar velocities of galaxies, the deviations of their motions from the homogeneous ``Hubble Flow" resulting from the inhomogeneous clustering of matter,  can give invaluable insights into the nature of gravity and the distribution of matter on cosmological scales. 

The growth of large scale structure (LSS) arises because of the infall of matter towards local overdensities. The peculiar velocity field of galaxies \cite{Kaiser:1987qv, 1995PhR...261..271S,Howlett:2022len} is a powerful tracer of the distribution of mass and its evolution over time \cite{Percival:2008sh,Hudson:2012gt,Said:2020epb,Turner:2024blz}. In recent years, the cosmological importance of velocity observables has grown substantially, driven in part by advances in measurements based on the kinematic Sunyaev-Zel'dovich (kSZ) effect and related LSS tracers \cite{Sunyaev:1980nv,Hand:2012ui,AtacamaCosmologyTelescope:2020wtv,Lague:2024czc,Gong:2025ffw}. Peculiar velocity and momentum field analyses have demonstrated that velocity-based observables contain information complementary to galaxy clustering alone for constraining the growth of structure and testing gravitational dynamics \cite{Springob:2014qja,Bullock:2017xww,Battaglia:2017neq}. 

Reconstruction of velocities from the density distribution, using the linear continuity equation, has been shown to be a valuable technique in multiple contexts,  including BAO reconstruction (e.g. \cite{Eisenstein:2006nj,Eisenstein:2006nk,Padmanabhan:2012hf,Xu:2012fw, BOSS:2013kxl,eBOSS:2020hur}) including in the presence of photometric redshifts \cite{Chan:2023zep}, the
characterization spectroscopic redshift space distortions (e.g. \cite{White:2015eaa,Chen:2019lpf}) and kinematic Sunyaev Zel'dovich effect detection \cite{ACTPol:2015teu}.


Improving the accuracy of cosmological velocity reconstruction is timely due to the increasing statistical power of current and upcoming LSS surveys \cite{Zhang_2013,Blake:2024lrh,Lai:2025xkf}.  
This includes ground-based instruments, such as the spectroscopic Dark
Energy Spectroscopic Instrument (DESI) \cite{DESI:2016fyo} and photometric Rubin Observatory
Legacy Survey of Space and Time (LSST) \cite{lsstsciencecollaboration2009lsstsciencebookversion}, and the Euclid \cite{Euclid:2024yrr} and upcoming Nancy Roman Space Telescopes \cite{Spergel:2015sza} which have both photometry and slitless spectroscopy.

Recently, a velocity reconstruction pipeline was developed and validated for Stage IV surveys using simulated galaxy catalogs that incorporated  observational realism (survey geometry, redshift-space distortions (RSD), and galaxy tracer selection) modeled for the DESI survey \cite{Guachalla:2023lbx, Hadzhiyska:2023nig}.
The velocity field was estimated from the galaxy density distribution using an approach similar to baryon acoustic oscillations (BAO) reconstruction; it adopts the Zel’dovich approximation at first order and uses the redshift-space continuity equation to map the density field into large-scale velocity estimates \cite{1970A&A.....5...84Z,Eisenstein:2006nk,Xu:2012hg}. 


In this work, we reconstruct the velocities more fully, by predicting the nonlinear residuals, the difference between the linear reconstructed and the true velocities.   Our goal is not to replace the standard reconstruction framework, but to augment it. We adopt the linear velocity reconstruction method of \cite{Guachalla:2023lbx, Hadzhiyska:2023nig} as our baseline estimator, and treat its output as a physically motivated first step reconstruction of the peculiar velocity. The intent is to characterize the nonlinear evolution as well as the effects on reconstruction of redshift-space distortions, survey geometry, redshift uncertainties, intrinsic tracer properties, and the smoothing required by the reconstruction itself. These effects 
induce structure in the residuals 
which are important for downstream analysis and rely not only on a strong correlation with the true velocity but also on an accurate calibration of its amplitude. Specifically, any residual bias in the reconstructed velocity can propagate into biased inference of cosmological constraints derived from velocity-based observables as well as kSZ amplitudes and cluster optical depths \cite{Ma:2017ybt,Turner:2023qau,Tishue:2025cvp}.

Machine learning methods are a natural tool for this task because they can learn nonlinear, nonlocal, and environment dependent mappings directly from simulations, e.g. \cite{Suresh:2024dsg,Gong:2025gmc}. Recent studies have shown that machine learning can be used to predict cosmic velocity fields when trained on dark matter halos.  U-Net \cite{2015arXiv150504597R} based approaches have been developed  that map gridded halo distribution or density fields to peculiar velocity fields in a periodic box setting \citep{Wang:2024ggp,Xiao:2024gcz}. We extend on these earlier machine learning studies, as well as linear reconstruction pipelines, in both methodology and scope. We move toward a more realistic observational setting. Rather than focusing on dark matter halos in periodic boxes, we study observable galaxy tracers in realistic lightcone catalogs that are more directly relevant for survey analyses. In addition, instead of relying on convolution-based architectures, we investigate a Transformer-based model as a more powerful deep learning framework \cite{2020arXiv201011929D,2021arXiv210314030L} for capturing nonlocal and multiscale correlations. These differences move the problem beyond proof of concept halo-level demonstrations and toward a reconstruction framework that is more directly applicable to real survey data analysis. 

 We use machine learning models, trained on simulations, to learn the residual between this baseline reconstruction and the true peculiar velocity. Conceptually, our method preserves the interpretability and robustness of the standard reconstruction on large scales, while allowing the machine learning model to absorb the nonlinear residuals that the linear theory doesn't recover.
By learning the systematic discrepancy between the first-order reconstructed velocity and the true velocity, we aim to improve both the fidelity of the reconstructed field and also its usefulness for downstream applications such as kSZ measurements and other probes sensitive to peculiar motions \cite{AtacamaCosmologyTelescope:2020wtv}.

We employ two machine learning models: a Gradient Boosting Decision Tree (GBDT), well suited to capture nonlinear relationships in tabulated features, and a Transformer architecture, designed to learn more complex, nonlocal feature interactions. Both models have been widely applied in astronomy (e.g., \cite{Li_2021, Sahakyan:2022uvr, Tolamatti:2023hef, Coronado-Blazquez:2023bbu, Donoso_Oliva_2023, Leung_2023, Gong:2024elx, Euclid:2024fdu}). The rationale for using these two machine learning models is that they represent different levels of modeling complexity and computational cost. We employ the GBDT as a simple and direct baseline, since this method only requires a set of input features and has been shown to perform strongly on structured learning tasks \cite{Friedman2001,Ke2017}. The Transformer architecture \cite{Vaswani2017} serves as a more advanced model. In contrast to GBDT, the Transformer offers much greater expressive power and flexibility, but it requires additional architectural design and is substantially more expensive to train. In this sense, GBDT serves as a fast and robust baseline, while the Transformer provides a more sophisticated model with the potential for additional improvement.

We assess the performance of the GBDT and Transformer models using luminous red galaxy (LRG) and emission line galaxy (ELG) samples modeled on DESI but also relevant to data expected from LSST, Euclid, and Roman. We test whether this method can improve line-of-sight (LOS) peculiar velocity reconstruction in both simulated periodic boxes and more realistic lightcone scenarios. Their performance is quantified with metrics in real space and Fourier space, including correlation based measures and statistics derived from the velocity power spectrum. We examine the stability of these improvements over two redshift bins of interest for ongoing and upcoming large-scale structure surveys. 

The paper is organized as follows: Section \ref{sec:data} describes the data used in this work. Section \ref{sec:methods} discusses the formalism of the baseline velocity reconstruction methods. Section \ref{sec:ML} describes the machine learning algorithms. The main velocity reconstruction results are presented in Section \ref{sec:results} and we discuss the applications of the reconstructed velocities in relation to kSZ analyses in Section \ref{sec:kSZ_results}. Finally, Section \ref{sec:conclusion} provides a summary of the key findings and possible avenues for future work.

\section{Data}
\label{sec:data}
We use simulation products from the AbacusSummit suite \cite{Maksimova:2021ynf}, generated with the Abacus $N$-body code \cite{Garrison:2018juw,Garrison:2021lfa}. AbacusSummit was developed for large-scale structure applications, with simulation volume, mass resolution, and numerical accuracy sufficient for DESI-era cosmological analyses \cite{DESI:2016fyo}. 

We use two complementary classes of data products: the simulation outputs themselves, which provide the true velocity field, and mock tracer catalogs of dark matter halos and galaxies,  constructed from those outputs.
 
Specifically we use the AbacusSummit\_base\_c000\_ph002 realization, which belongs to the base resolution set and adopts the fiducial c000 cosmology with phase label ph002. This simulation evolves $6912^3$ particles in a periodic box of side length $L_{\rm box}=2000,h^{-1}{\rm Mpc}$, corresponding to a particle mass of $m_{\rm p}\simeq 2.1\times10^9,h^{-1}M_\odot$. The fiducial c000 cosmology is a flat $\Lambda$CDM model that follows the Planck 2018 baseline cosmology \cite{Planck:2018vyg}. The corresponding cosmological parameters are $\Omega_b h^2=0.022$, $\Omega_{\rm cdm} h^2=0.12$, $h=0.67$, $A_s=2.08\times10^{-9}$, and $n_s=0.97$.

We focus on LRG and ELG tracer samples relevant for DESI analysis. We construct DESI-like mock catalogs using the AbacusHOD framework \cite{Yuan:2021izi,Yuan:2023ezi}, an extended halo occupation distribution (HOD) model that populates dark matter halos with galaxies. The standard HOD model of \cite{Zheng:2007zg} is used with model parameter values determined in \cite{Hadzhiyska:2023fic} using the
DESI SV3 (Survey Validation 3) data. Since the reconstruction is sensitive to the galaxy bias, performing better for higher bias samples, we follow \citep{Hadzhiyska:2023nig} and have a higher minimum mass cut for the lower bias ELG sample, relative to the higher bias LRG sample, increasing the HOD logarithmic mass cut parameter by 0.5 dex. The HOD model populates dark matter halos with central and satellite galaxies and incorporates linear velocity biases to the central galaxy relative to the halo velocity based on the halo velocity dispersion and separate velocity biases for each satellite galaxy relative to the velocity of relevant dark matter particle. The model parameters are obtained by calibration to the correlation function on scales below $\sim$30 Mpc/$h$ from the DESI Survey Validation 3 (SV3) data at two pivot redshifts, $z=0.5$ and $z=0.8$, for the LRG sample and two, $z=0.8$ and $z=1.1$, for the ELG sample. Following the bias modeling in \cite{Hadzhiyska:2023nig} based on \cite{2010ApJ...724..878T}, we assume $\bar{b}_{LRG}(z)\approx 2.2$ for both the LRG samples centered at $z=0.5$ and $z=0.8$, and with $\bar{b}_{ELG}\approx 1.3$ and 1.6 for the ELG samples centered at $z=0.8$ and 1.1.

We consider two complementary realizations of the tracer sample: a periodic box catalog and a lightcone catalog. In the box case, we focus only on a box snapshot at $z=0.5$ for the LRG sample. Galaxies are populated with AbacusHOD on fixed-redshift periodic snapshots from the AbacusSummit simulation. For the lightcone catalogs, we consider two redshift bins of DESI-like LRG samples, $0.4<z<0.6$ and $0.7<z<0.9$, centered at $z=0.5$ and $z=0.8$, respectively. We also study the $0.7<z<0.9$ and $1.0<z<1.2$ redshift bins of DESI-like ELG samples. Unlike the periodic box, galaxies are instead populated on the AbacusSummit halo lightcone catalogs \cite{Hadzhiyska:2021uhm}, which were specifically designed for efficient mock generation with AbacusHOD. This lightcone realization incorporates the redshift evolution of the tracer population together with the survey geometry relevant for realistic DESI-like observations. 

We summarize the mock galaxy sample characteristics in Table~\ref{tab:galnum}. 
The comoving number densities are consistent with the DESI target densities, with $n(z)\sim5\times 10^{-4}h^3$/Mpc$^3$ and $1\times 10^{-3}h^3$/Mpc$^3$ for the Main and Extended LRG samples ($0.4<z<0.8$) and $n(z)\sim5\times 10^{-4}h^3$/Mpc$^3$ for the ELG sample ($0.85<z<1.2$).

\begin{table}
\centering
\setlength{\tabcolsep}{6pt}
\renewcommand{\arraystretch}{1.25}
\begin{tabular}{|c|c|c|c|c|c|}
\hline
Geometry & Tracer & Redshift & $N_{gal}$ & $n(z)$ & $\bar{b}$
\\
\hline 
Box & LRG & $0.4 < z < 0.6$ & 5,063,029 & 0.63 & 2.2 \\
\hline
\multirow{4}{*}{Lightcone} 
& \multirow{2}{*}{LRG} 
& $0.4 < z < 0.6$ & 750,121 & 0.61 & 2.2\\
&  
& $0.7 < z < 0.9$ & 1,533,969 & 0.69 & 2.2\\
\cline{2-6}
& \multirow{2}{*}{ELG} 
& $0.7 < z < 0.9$ & 636,018 & 0.29 & 1.3\\
&  
& $1.0 < z < 1.2$ & 472,278 & 0.16 & 1.6\\
\hline
\end{tabular}
\caption{Summary of the simulated galaxy samples used in this work. We list the simulation geometry, tracer type, redshift range, number of galaxies $N_{\rm gal}$,  comoving number density $n(z)$ ($[10^{-3}h^3\,\mathrm{Mpc}^{-3}]$) and assumed mean bias, $b$ for the mock catalogs.}
\label{tab:galnum}
\end{table}

\section{FORMALISM}
\label{sec:methods}

\subsection{Velocity Reconstruction}
\label{sec:reconstruction}
Following \cite{Guachalla:2023lbx, Hadzhiyska:2023nig}, we reconstruct the peculiar velocity from the observed galaxy overdensity. At linear order, the redshift-space galaxy density contrast is related to the velocity field through the continuity equation. Assuming linear galaxy bias \cite{Desjacques:2016bnm}, $\delta_g = b\,\delta_m$, the redshift-space relation can be written as
\begin{equation}
\nabla \cdot {\bf v} + f \, \nabla \cdot \left[ (\bf{v}\cdot \hat{{n}})\hat{{n}} \right]
= - a H f \, \frac{\delta_g}{b},
\label{eq:continuity_redshift}
\end{equation}
where $H$ is the Hubble parameter, $a=1/(1+z)$ is the scale factor, $f \equiv d\ln D / d\ln a$ is the linear growth rate, and $\hat{{n}}$ denotes the LOS direction. Rather than solving Eq.(\ref{eq:continuity_redshift}) directly for the velocity, it is convenient to first infer the large-scale displacement field, ${\Psi}$, following the approach used for BAO reconstruction \cite{Eisenstein:2006nk,Xu:2012hg}, and then convert the displacement field to velocity.

We begin by assigning galaxies to a three-dimensional mesh and smoothing the tracer overdensity with a Gaussian kernel,
\begin{equation}
W_G(k) = \exp\!\left(-\frac{k^2 R_s^2}{2}\right),
\label{eq:gaussian_smoothing}
\end{equation}
where $R_s$ is the comoving smoothing scale. This smoothing suppresses nonlinear and shot noise dominated modes, so that the subsequent reconstruction is driven primarily by the linear large-scale density field.

In linear theory, the reconstructed displacement in Fourier space can be written as
\begin{equation}
{\Psi}({k})
=
\frac{i{k}}{k^2}
\frac{W_G(k)}
{b + f \mu^2}\delta_g^{\,s}({k}),
\qquad
\mu \equiv \hat{{k}}\cdot \hat{{n}} .
\label{eq:psi_fourier}
\end{equation}
This expression corresponds to the inverse gradient solution of the linearized continuity equation and provides the large-scale Zel'dovich displacement inferred from the observed redshift-space galaxy field. 

In practice, we reconstruct the displacement field numerically using the MultiGrid  implementation \cite{White:2015eaa} provided by the \textsc{pyrecon}\footnote{\url{https://github.com/cosmodesi/pyrecon}} package for both the box and lightcone catalogs. For the box catalogs, we assume a periodic cubic volume and adopt a fixed LOS direction,  $\hat{\bm z}$. For the lightcone catalogs, we first convert the galaxy catalogs from $(\mathrm{RA},\mathrm{DEC},z)$ to Cartesian comoving coordinates, ${\bf x}=\chi {\bf \hat{n}}$, and then perform the reconstruction in three dimensional space with a local, position dependent LOS direction. 

The reconstructed displacement field is then evaluated at the positions of the target objects. The reconstructed peculiar velocity is then obtained via,
\begin{equation}
 {\bf v}_{linear}(\bf{x}) = a H(a) f(a) \, {\bf \Psi}(\bf{x}).
\label{eq:vrec_from_psi}
\end{equation}
To extend beyond the linear estimate we model the nonlinear velocity by using the machine learning algorithms described in Section \ref{sec:ML} to predict the residual velocities, $\Delta {\bf v}$,
\begin{equation}
 \Delta{\bf v}(\bf{x}) ={\bf v}_{{\rm true}}(\bf{x})-{\bf v}_{{\rm linear}}(\bf{x}).
 \label{eq:residual_target}
\end{equation}
In many applications, especially kSZ measurements, the LOS peculiar velocity component, $v_{\parallel}$, is the main quantity of interest, although the transverse component, $v_{\perp}$, also provides valuable information, for example in the moving lens effect \cite{Hotinli:2021hih, Hotinli:2023ywh}:
\bea
v_{\parallel}&=&{\bf v}\cdot \hat{{n}},
\\
v_{\perp}&=&{\bf v}\cdot \hat{e}_{\perp},
\eea
with,
\bea
\hat{e}_{\perp} = \frac{(\hat{n}_z,\;\hat{n}_z,\;-\hat{n}_x-\hat{n}_y)} {\sqrt{2\hat{n}_z^2+(\hat{n}_x+\hat{n}_y)^2}}.
\eea
We focus on reconstructing the  halo velocities from the galaxy samples, rather than reconstructing the galaxy velocities themselves \cite{Guachalla:2023lbx}, as the halo bulk motion governs the kSZ signal \citep{Soergel:2017ahb,Hadzhiyska:2023cjj}.

As a characterization of the reconstruction, we consider the power spectrum of the LOS momentum field. For a galaxy sample, we construct the LOS momentum field from the galaxy overdensity and velocity fields as
\begin{equation}
q_{\parallel}({\bf x}) = [1+\delta_g({\bf x})]\, v_{\parallel}({\bf x}),
\label{eq:momentum_field}
\end{equation}
where $\delta_g(\bf{x})$ denotes the galaxy overdensity field and $v_{\parallel}(\bf{x})$ is the LOS peculiar velocity field. Its two-point statistics in Fourier space define the momentum power spectrum,
\begin{equation}
\langle q_{\parallel}({\bf k}) q_{\parallel}({\bf k}') \rangle
=
(2\pi)^3 P_{q_{\parallel}q_{\parallel}}(k)\,\delta_D({\bf k}-{\bf k}').
\label{eq:momentum_pk}
\end{equation}

We use $P_{q_{\parallel}q_{\parallel}}(k)$ as an additional diagnostic of the reconstruction. Because it depends on both the density weighting and the velocity field, it provides a useful way to track the scale dependence of the reconstructed signal.

\subsection{Reconstruction Efficacy Metrics}
\label{subsec:metric}
To quantify reconstruction performance, we consider both object-level and power-spectrum-level metrics.

For a set of objects with true  velocities ($v_{\parallel}$ or $v_{\perp}$), $v_{{true},i}$ and reconstructed counterparts, $v_{{rec},i}$, we first compute the Pearson correlation coefficient, measuring the strength of the linear association between $v_{\rm true}$ and $v_{\rm rec}$,
\begin{equation}
r_v \equiv
\frac{\sum_i \left(v_{{true},i}-\bar{v}_{true}\right)\left(v_{{rec},i}-\bar{v}_{rec}\right)}
{\sqrt{\sum_i \left(v_{{true},i}-\bar{v}_{true}\right)^2}
 \sqrt{\sum_i \left(v_{{rec},i}-\bar{v}_{rec}\right)^2}},
\label{eq:pearson_v}
\end{equation}
where $\bar{v}_{true}$ and $\bar{v}_{rec}$ are the sample means of the true and reconstructed $v_{\parallel}$ or $v_{\perp}$ components.  

Since $r_v$  is insensitive to an overall bias in amplitude or mean offset, we also use the concordance correlation coefficient (CCC), defined as
\begin{equation}
\rho_v \equiv
\frac{2\,{cov}(v_{true},v_{rec})}
{\sigma_{true}^2 + \sigma_{rec}^2 + \left(\bar{v}_{true}-\bar{v}_{rec}\right)^2},
\label{eq:ccc_v}
\end{equation}
where $\sigma_{true}^2$ and $\sigma_{rec}^2$ denote the sample variances and ${cov}(v_{true},v_{rec})$ is the sample covariance. Unlike the Pearson correlation coefficient, $\rho_v$ penalizes not only random scatter but also differences in the overall offset and amplitude. Therefore, this provides a stricter measure of agreement between the reconstructed and true velocities.

To assess the scale dependence of the reconstruction, we further compute the correlation coefficient of the LOS momentum power spectrum in Fourier space. Let $q_{true}$ and $q_{rec}$ denote the true and reconstructed LOS momentum fields, we define their auto- and cross-power spectra through
\begin{equation}
\langle q_a({\bf k}) q_b({\bf k}') \rangle
\equiv
(2\pi)^3 P_{q_a q_b}(k)\,\delta_D({\bf k}-{\bf k}'),
\label{eq:momentum_crosspk}
\end{equation}
with $a,b \in \{{true,rec}\}$.
The corresponding scale dependent cross-correlation coefficient is, 
\begin{equation}
r_q(k) \equiv
\frac{P_{q_{true} q_{rec}}(k)}
{\sqrt{P_{q_{true} q_{true}}(k)\,P_{q_{rec} q_{rec}}(k)}}.
\label{eq:rqk}
\end{equation}
with $r_q(k)<1$ indicating a loss of correlation induced by an imperfect reconstruction.
\subsection{Connections to kSZ analyses}
As part of this work, we also consider how the reconstructed velocities enhance two common kSZ analyses: kSZ pairwise correlations and stacked kSZ profile estimation.  

\subsubsection{Pairwise Velocity Statistic}
\label{subsec:pairwise}

The pairwise velocity is a useful summary statistic of the cosmic velocity field,  characterizing the average relative in-fall of tracer pairs as a function of their comoving separation. Following \citep{Ferreira:1998id}, the pairwise velocity estimator in a separation bin centered at $r$ is,
\begin{equation}
V(r)=
\frac{\sum_{i<j}
\left(v_{\parallel,i}-v_{\parallel,j}\right)c_{ij}}
{\sum_{i<j} c_{ij}^{\,2}}.
\label{eq:pairwise_estimator}
\end{equation}
The sum is over all tracer pairs whose comoving separation, $r=|r_i-r_j|$, falls within the separation bin and $c_{ij}$ is the geometric projection factor,
\begin{equation}
c_{ij} \equiv \hat{r}_{ij}\cdot
\frac{\hat{n}_{i}+\hat{n}_{j}}{2} =\frac{(r_i-r_j)(1+\cos\alpha)}
{2\sqrt{r_i^2+r_j^2-2r_ir_j\cos\alpha}},
\label{eq:cij1}
\end{equation}
where $\hat{r}_{ij}$ is the unit vector along their separation and $\alpha$ is their angular separation.

The kSZ is a secondary  anisotropy imprinted on the primary CMB due to a Doppler shift due to line of sight peculiar motion of baryonic matter it interacts with. For an isolated galaxy cluster, for example, one might model the kSZ temperature as
\begin{equation}
\frac{\Delta T_{kSZ,i}(\theta)
}{T_{0}}=-\tau_i(\theta)\,
\frac{v_{\parallel,i}}{c},
\label{eq:ksz_halo}
\end{equation}
where  $T_{0}=2.726K$ is the  primary CMB temperature, $c$ is the speed of light, $\tau_i$ is the projected cluster optical depth and $\theta$ is the angular distance from the halo center.

The pairwise kSZ momentum and velocity pairwise correlations are related by:
\begin{equation} \label{eq:PVtau}
 P_{kSZ}(r) = -  \frac{T_0}{c} \bar{\tau}_{eff} V(r)
\end{equation} 
where $\bar{\tau}_{eff}$ is an effective mass-averaged optical depth across the cluster sample as a whole.

In previous work \cite{Gong:2025gmc}, we showed how cluster optical depth estimates obtained from machine learning modeling could be combined with kSZ temperature measurements to obtain unbiased estimates of the cluster pairwise velocity correlation. 

Here we consider an alternative application of the kSZ pairwise measurements, in which we combine the pairwise velocity correlation, $V$, obtained from galaxy velocity reconstruction with pairwise kSZ momentum measurements to estimate an effective average optical depth for a cluster sample, $\tau_{eff}$. 
 
We create a simulated kSZ map based on the \texttt{AbacusSummit} halo catalog and add primary CMB contamination, beam effects and instrument noise based on Simons Observatory survey specifications. Details of the kSZ map simulation and pairwise kSZ momentum optical depth fitting are respectively provided in Appendices ~\ref{subsec:kSZ_map} and ~\ref{subsec:AP}. 

\subsubsection{Stacked kSZ profile}
\label{subsec:kSZ_profile}
We also consider the application of  the reconstructed velocities to estimate a kSZ-derived halo electron density profile, a useful probe of the spatial distribution of ionized gas in galaxy clusters. This is obtained by stacking the kSZ temperatures centered on cluster halos after weighting by their LOS velocities. In our analysis we apply the reconstructed velocities to simulated kSZ maps to directly assess their utility for accurately estimating the kSZ-derived halo profile.

We construct a semi-analytic kSZ map using the procedure described in Appendix~\ref{subsec:kSZ_map}, and extract the temperature around each halo using the aperture photometry filter described in Appendix~\ref{subsec:AP}. In addition to the AP-filtered temperature, $T_{AP}$, we also consider the average disk temperature, $T_{Disk}$, to compare two commonly used choices for measuring the kSZ signal. We then examine whether the reconstructed velocities can recover the stacked kSZ profile defined in Eq.~(\ref{eq:kSZ_profile}), using the LRG $0.4 < z < 0.6$ redshift sample.

To quantify the average kSZ signal recovered by different velocities, we construct a stacked kSZ profile using a velocity weighted estimator, following \citep{AtacamaCosmologyTelescope:2020wtv}. For a given aperture radius $\theta_d$, let $T_i(\theta_d)$ denote the filtered kSZ temperature measured around tracer $i$. In practice, $T_i(\theta_d)$ can be taken to be either the aperture photometry temperature $T_{AP,i}(\theta_d)$ or the disk averaged temperature $T_{Disk,i}(\theta_d)$. We then calculate the stacked velocity-weighted kSZ signal as \cite{AtacamaCosmologyTelescope:2020wtv}
\begin{equation}
\hat{T}_{kSZ}(\theta_d)
=
-\frac{1}{r_{corr}}\frac{\sigma_v^{\parallel}}{c}
\frac{
\sum_{i=1}^{N_{tr}} T_i(\theta_d)\left(v_{\parallel,i}/c\right)
}{
\sum_{i=1}^{N_{tr}} \left(v_{\parallel,i}/c\right)^2
},
\label{eq:kSZ_profile}
\end{equation}
where $r_{corr}$ is the correlation coefficient, 
$\sigma_v^{\parallel}$ is the root-mean-squared of the reconstructed velocities, and $N_{tr}$ is the number of tracers.

We compare the performance of the approach using Eq.~(\ref{eq:kSZ_profile}) in recovering the halo density profile for temperatures measured using $T_{AP}$ and $T_{Disk}$ and different choices for the input velocity field: the true halo velocity, the linear velocity approximation, and the machine learning derived velocity. 

\section{Machine learning}
\label{sec:ML}
In Section \ref{sec:feature}, we describe the features used in the machine learning models. Section \ref{sec:train} outlines the general training and validation process for the models. Sections \ref{sec:GBDT} and \ref{sec:Transformer} introduce the two principal machine learning models used in this work: GBDT and Transformer.

\subsection{Model Features}
\label{sec:feature}
Our machine learning models  are designed to improve upon the linearly reconstructed LOS velocity by using the velocity residual, defined in (\ref{eq:residual_target}), as the regression target inferred from its relationship with a suite of model features as might be derived from the 3D position and density maps measured in spectroscopic or photometric galaxy surveys.

The model features include the set of basic global parameters ($\hat{n}$, z), plus three observable-derived groups:  multi-scale displacement (location-derived) features, multi-scale environment (density-derived) features and cross-scale differences in both displacement and environment features.  

Following \cite{Guachalla:2023lbx}, each of the input features are constructed at three smoothing scales, $R_s$ =10.5,\ 12.5,\ 14.8 $h^{-1}{\rm Mpc}$. The final velocity prediction is obtained by adding the predicted residual back to the linear  reconstructed baseline, smoothed with a scale of \(R_s=12.5\,h^{-1}{\rm Mpc}\).

For any feature \(q\) available at multiple smoothing scales, the cross-scale difference features use the \(R_s=12.5\,h^{-1}{\rm Mpc}\) reconstruction as the reference, 
\begin{equation}
\Delta q(R_s)=q(R_s)-q(12.5 h^{-1}{\rm Mpc}).
\end{equation}

In machine learning modeling, difference features are frequently grouped or treated separately from the original multi-scale features because they encode relationships between variables rather than the variables themselves. In that sense, the difference features here are not independent of the displacement or environment features but they carry a different semantic meaning: they describe scale-dependence rather than the absolute value at one scale.

For each galaxy and for each smoothing scale, we use a core set of displacement features that provides the baseline information. These include the reconstructed displacement field 
$\bm{\Psi}({\bf x})$, its magnitude, $|\Psi|$,  its LOS projection, $\Psi_\parallel = \bm{\Psi}\cdot \hat{\bm n},$ and the linearly reconstructed velocity, $aH(a)f(a)\,\bm{\Psi}\cdot\hat{\bm n}$, which directly correspond to the standard first-order reconstruction of the peculiar velocity from the continuity equation. 

We also consider an extended set of nonlinear and dimensional displacement features: $\Psi_{\perp}=\left[|\bm{\Psi}|^2
-(\bm{\Psi}\cdot \hat{\bm{n}})^2
\right]^{1/2},$   $\Psi_\parallel^3$, $\Psi_\parallel|\Psi_\parallel|$,
$\Psi_\parallel/|\bm{\Psi}|,
|\bm{\Psi}|/R_s$, 
and $|\Psi_\parallel|/R_s$.

The environment features characterize the local and anisotropic structure around each tracer, evaluated at the galaxy position for each smoothing scale. These density-dependent features 
complement the displacement features by providing information about the local structure in which the velocity field is embedded. A core set of environment features provide scalar and vector density descriptors that encode the local density field and the direction and steepness of local density variations: the smoothed overdensity, based on galaxy number counts, $\delta_g$, and its Laplacian, $\nabla \delta_g$, and their magnitudes. An extended set of environment features include tensor descriptors, the Hessian, $H_{ij}\equiv \partial_i\partial_j\delta_g,$ and Laplacian, $\nabla^2\delta_g$, and scalar descriptors of the tidal environment derived from the traceless tidal tensor
\begin{equation}
T_{ij}
\equiv
\partial_i\partial_j\delta_g
-\frac{1}{3}\delta_{ij}\,\nabla^2\delta_g.
\end{equation}
From this tensor, we define the shear amplitude as
$\sqrt{\sum_{i,j} T_{ij}T_{ij}}$,
its squared amplitude,
$\sum_{i,j} T_{ij}T_{ij}$,
and an anisotropy measure,
$\sqrt{\sum_{i,j} T_{ij}T_{ij}}/(|\delta_g|+\epsilon)$,
where \(\epsilon\) is a small positive constant introduced for numerical stability. The tidal tensors describe the anisotropic curvature and shear, and cosmic web environment.  Similar density, tidal, and Hessian-based quantities have been widely used in perturbative bias modeling, cosmic web classifications, and machine learning studies of structure formation, where features beyond the scalar density are needed to characterize environment dependent evolution \cite{Forero-Romero:2008svv,Desjacques:2016bnm,Lucie-Smith:2019hdl}. Together, these feature categories allow the models to combine multiscale reconstruction information while learning scale-dependent corrections to the linear LOS velocity estimate.

The environment features may be noisier than the displacement features because quantities such as $\nabla\delta_g$, $H_{ij}$, $\nabla^2\delta_g$, and the tidal tensor descriptors, are constructed from spatial derivatives of the reconstructed density field, which can amplify the impact of survey geometry, sparse sampling, redshift space distortions, and redshift uncertainties in lightcone data.  

We therefore compare the reconstruction performance of three feature sets with  different levels of complexity.
The {\it ``full"} feature set including all displacement and environment features is analyzed for both the cubic box and lightcone analyses. For the lightcone two more conservative subsets are also considered: a {\it ``displacement-only"} set to focus on the features principally derived from spectroscopic data and a {\it ``core-only"} set just using the core, principally scale and vector, displacement and environment features. Each set includes both the smoothed and corresponding cross-scale difference features. These comparisons allow us to quantify how much improvement comes from robust displacement information alone, how much is retained after adding only the simplest density gradient information, and how much additional information is contributed by the more complex but potentially noisier tensor derived environment features.
 
\subsection{Train and Test Process}
\label{sec:train}
To train and evaluate the machine learning models, we split the full sample into training and testing subsets with a 7/3 ratio. The training set is used for model fitting and hyperparameter selection, while the testing set is kept aside and used only for the final evaluation of the trained model on unseen data.

To reduce overfitting and improve the robustness of the model, we perform cross validation on the training set. In particular, the training sample is divided into five folds, and in each iteration the model is trained on four of the folds and validated on the remaining fold. The cross validation performance is used to determine the stopping point of the training and to monitor the generalization behavior of the model.

We optimize the model hyperparameters with Optuna\footnote{\url{https://optuna.org/}}. For each trial, a set of hyperparameters is proposed, the model is trained on the training folds, and its performance is evaluated through cross validation. The hyperparameter set that yields the best validation performance is then adopted for the final model. 
For the training objective, we adopt the mean squared error (MSE),
\begin{equation}
\mathrm{MSE}
=
\frac{1}{N}
\sum_{i=1}^{N}
\left(
\hat{\Delta v}_{i}
-
\Delta v_{i}
\right)^2,
\label{eq:mse}
\end{equation}
where $\Delta v_{i}$ is the target residual (LOS or transverse) velocity for galaxy $i$, and $\hat{\Delta v}_{i}$ is the corresponding model prediction. After the stopping criterion and hyperparameters are determined, the model is retrained on the full training sample and its performance is finally assessed on the independent test sample.

For both the GBDT and Transformer models, the final reconstructed  LOS or transverse velocity estimate is obtained by adding the predicted residual correction to the linear reconstruction velocity,
\begin{equation}
v_{\mathrm{GBDT/Transformer}}
=
v_{\mathrm{linear}}
+
\Delta v_{\mathrm{pred}}.
\label{eq:vfinal}
\end{equation}

\subsection{GBDT}
\label{sec:GBDT}
For our first machine learning model, we use a gradient boosting decision tree (GBDT) regressor implemented with LightGBM\footnote{\url{https://lightgbm.readthedocs.io/en/stable/}}. GBDT builds an ensemble of decision trees sequentially, with each new tree trained to reduce the residual error of the previous ensemble. This model is well suited to tabular data and can naturally capture nonlinear feature interactions without requiring explicit feature engineering beyond the physically motivated variables introduced in Section~\ref{sec:feature}.

In our application, the GBDT model takes the concatenated multi-scale feature vector as input and predicts the residual LOS velocity defined in Eq.~(\ref{eq:residual_target}) with the MSE in Eq.~(\ref{eq:mse}) as the training objective. The final LOS velocity estimate is then obtained using Eq.~(\ref{eq:vfinal}). With its computational efficiency and strong performance on structured inputs, GBDT serves as a baseline model in our analysis.

We report the main hyperparameters used for model training. For the LightGBM model, L2 regularization coefficients were set to 0.01. The learning rate was set to 0.044. The model complexity was controlled using 256 leaves, a maximum tree depth of 10, and a minimum of 100 samples per leaf. We also applied bagging with a bagging fraction of 0.8 and a bagging frequency of 5 to improve generalization.

\begin{figure*}[t]
    \centering
    \includegraphics[width=\textwidth]{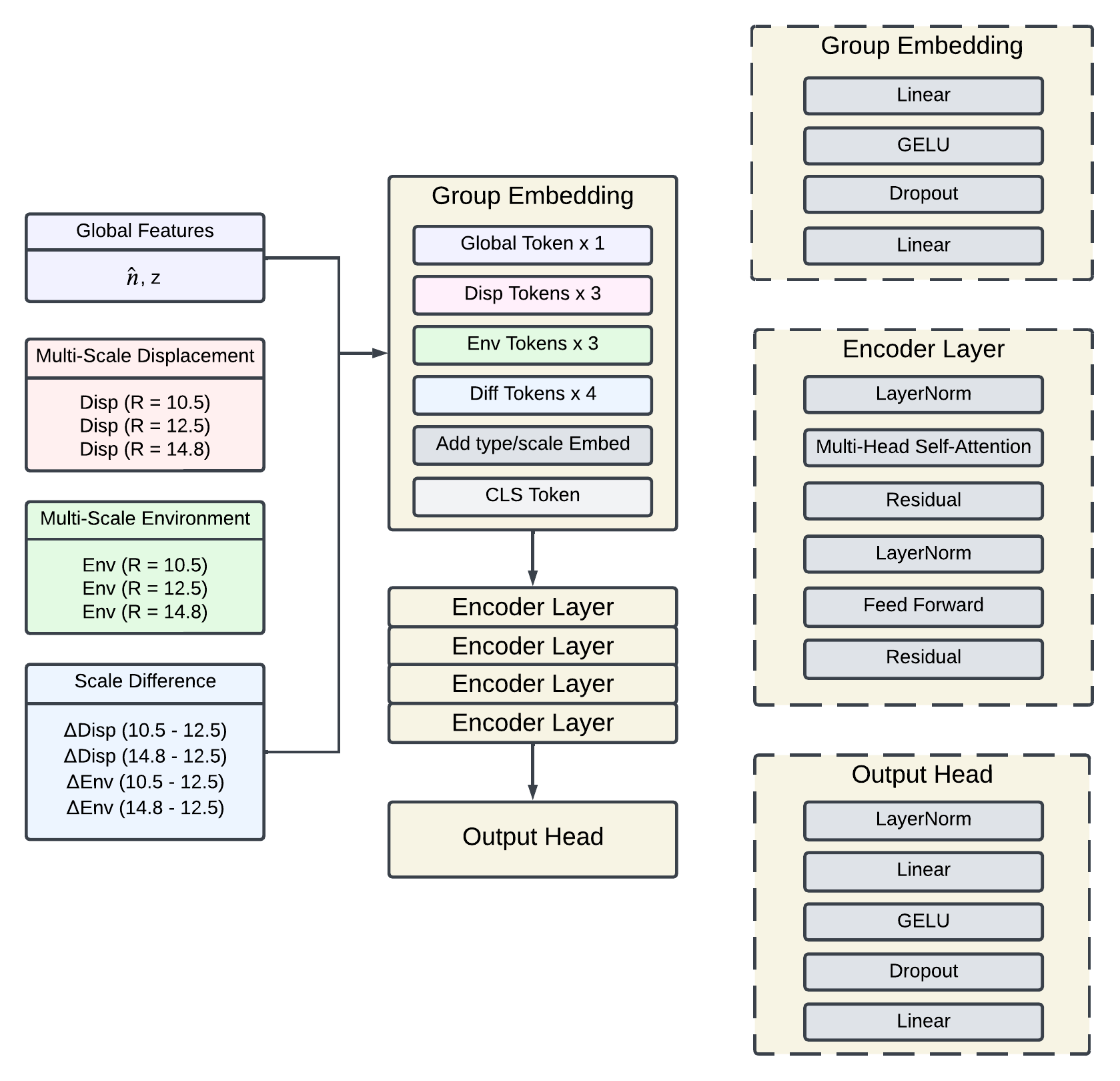}
    \caption{Illustration of the Transformer architecture used in this work. [Left] The overall workflow of our model. [Right] The detailed components of the group embedding, Transformer encoder layer, and output head. The terms used in the figure are discussed in Sec.\ref{sec:Transformer}}
    \label{fig:transformer_architecture}
\end{figure*}

\subsection{Transformer}
\label{sec:Transformer}
Transformer is a neural network architecture originally developed for sequence modeling, in which self-attention is used to learn how different elements of the input should interact with each other. In our application, the Transformer is used not to model long sequential structure, but to learn nonlinear couplings among reconstruction features derived from different smoothing scales and feature types.

\begin{figure*}[t!]
    \centering
    \includegraphics[width=0.86\textwidth]{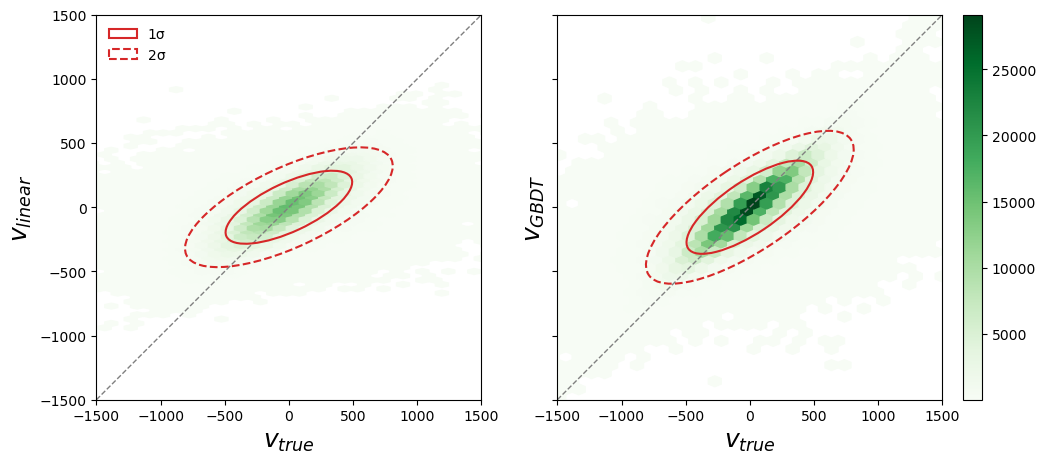}
    \caption{Comparison of the reconstructed and true LOS peculiar velocities in the periodic box for the LRG sample centered at redshift z = 0.5. [Left] shows the baseline linear reconstruction, while [Right] shows the GBDT residual corrected reconstruction. The dashed diagonal line shows perfect reconstruction 
    , and the contours indicate the $1\sigma$, $2\sigma$, and $3\sigma$ regions. The green background shows a hexagonal-binned density map of the distribution, where darker regions correspond to a higher number density of objects per redshift bin.
    }
    \label{fig:box_velocity}
\end{figure*}

Rather than treating all input variables as a single flattened vector, we partition the features as introduced in Section~\ref{sec:feature} in four semantically defined groups: global galaxy properties (e.g. directional information and redshift), reconstruction displacement features at different smoothing scales, environmental features at different smoothing scales, and difference features constructed relative to a reference smoothing scale. Each feature group is treated as one token, so that the input for each galaxy is represented as a sequence,
${\bf G} = \{\bm{g}_1,\bm{g}_2,\dots,\bm{g}_{N_G}\},$
where $\bm{g}_i$ denotes the feature vector of group $i$, and $N_G$ is the total number of feature groups.

Each token is first mapped into a latent space of fixed dimension through a learnable group-specific embedding network. To preserve the semantic identity of each token, we add learnable embeddings that encode both the feature type and the associated smoothing scale. The initial token representation is therefore written as
\begin{equation}
\bm{h}_i^{(0)} = f_i(\bm{g}_i) + \bm{e}^{\rm type}_i + \bm{e}^{\rm scale}_i,
\end{equation}
where $f_i$ denotes the embedding network for group $i$, while $\bm{e}^{\rm type}_i$ and $\bm{e}^{\rm scale}_i$ are the learnable type and scale embeddings, respectively. In addition, a learnable {\it classification} token ({\tt [CLS]}) is prepended to the input sequence to collect information from all feature groups.

The resulting token sequence is passed through a stack of Transformer encoder layers. Each encoder layer consists of a multi-head self-attention block followed by a feed forward network. Through self-attention, each feature group can interact with all others, allowing the model to learn which groups are most informative and how information from different smoothing scales and feature types should be combined. In this setting, the role of attention is to capture complementarity among multi-scale reconstruction features rather than long range positional dependence.

After the final encoder layer, we use the output representation of the {\tt [CLS]} token as a compact summary of the full input. This summary vector is then passed to a multilayer perceptron regression head to predict the residual LOS velocity,
\begin{equation}
\Delta v_{{pred}}
=
f_{\rm head}\!\left(\bm{h}_{\rm CLS}^{(L)}\right),
\end{equation}
where $\bm{h}_{\rm CLS}^{(L)}$ is the {\tt [CLS]} representation after the final encoder layer and $f_{\rm head}$ denotes the regression head. The architecture of the model is summarized in Fig. \ref{fig:transformer_architecture}. As shown in the figure, the $Linear$ layers are used to transform the raw feature vectors into trainable token representations with a unified dimensionality. The {\it Gaussian   Error   Linear  Unit} (GELU) activation layer introduces nonlinearity, enabling the model to capture more complex feature interactions beyond simple linear mappings. The {\it Dropout} layer is used to regularize the model and reduce overfitting. {\it LayerNorm} is applied to stabilize the feature distribution and improve training convergence. The {\it multi-head self-attention} layer allows each token to model global dependencies and interactions among different tokens. $Residual$ layer is introduced to preserve the original information and facilitate gradient propagation in deep networks. The {\it Feed Forward} layer then performs token-wise nonlinear transformation, further enhancing the expressive power of the learned representations.

We also report the main training hyperparameter of the Transformer model. The input features are embedded into 128 dimensional tokens. The Transformer encoder consists of 4 attention heads, 4 encoder layers, a feed-forward dimension of 256, and a dropout rate of 0.1. The model was trained using the AdamW optimizer with a learning rate of $1\times10^{-3}$, weight decay of $1\times10^{-4}$, a batch size of 4096, and mean squared error loss. A \texttt{ReduceLROnPlateau} scheduler was used to reduce the learning rate by a factor of 0.5 if the validation loss did not improve for 4 epochs. Gradients were clipped to a maximum norm of 1.0 to stabilize training.

\section{Results}
\label{sec:results}
In this section, we present the results of our machine learning models for velocity reconstruction in both box and lightcone settings. In Section. \ref{sec:box}, we first examine LOS velocity reconstruction in the cubic box case for the LRG sample at redshift z = 0.5, using the GBDT model. This provides a simple, computationally efficient benchmark for evaluating the improvement gained from residual learning. In Section~ \ref{sec:lightcone}, we discuss the performance of both the GBDT and Transformer machine learning models in the more realistic lightcone setting, considering both LOS and transverse velocity reconstruction for LRG and ELG samples in redshift ranges $0.4 < z < 0.6$, $0.7 < z < 0.9$, and $1.0 < z < 1.2$. We further investigate the impact of redshift uncertainties in Section. \ref{sec:Zerr}. 

\subsection{Box Reconstruction}
\label{sec:box}
We begin by evaluating the reconstruction performance in the periodic box setting. In this controlled setup, we compare the baseline linear theory reconstruction with the GBDT residual corrected prediction.

\begin{figure}[t!]
    \centering
\includegraphics[width=\columnwidth]{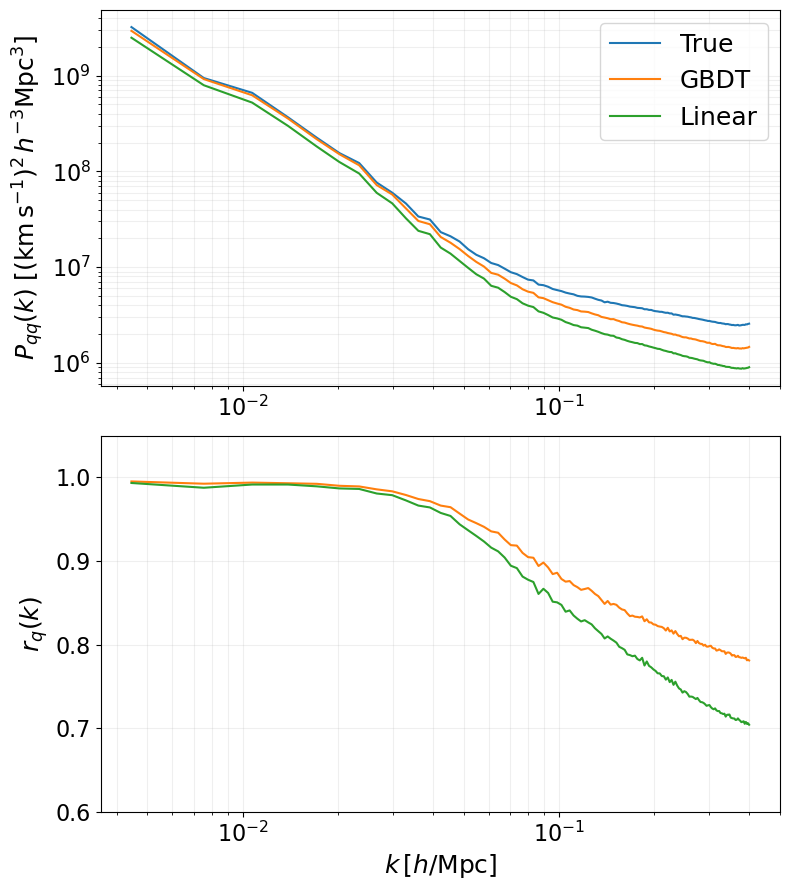}
    \caption{Comparison of the true [blue], linear reconstructed [green], and GBDT machine learning corrected [orange] LOS momentum fields in the periodic box for the LRG sample 0.4$<z<$0.6 for the momentum power spectrum $P_{qq}(k)$ [top] and the cross-correlation coefficient  $r_q(k)$ [bottom].
    }
    \label{fig:box_pscorr}
\end{figure}

Fig. \ref{fig:box_velocity} shows the reconstructed LOS velocities against the true values. The machine learning correction leads to a visibly tighter relation and better overall agreement with the true velocities. We find that the linear reconstruction can capture the overall trend of the velocity dispersion, but it exhibits a relatively broad scatter around the one-to-one relation and a visible suppression in the reconstructed amplitude. After applying the GBDT residual correction, the reconstructed velocities become substantially more concentrated around the diagonal, with a clear reduction in scatter and a better recovery of the true amplitude.

For the linear reconstruction baseline with smoothing scale \(R_s = 12.5\,h^{-1}\mathrm{Mpc}\), the Pearson correlation coefficient between the reconstructed and true velocities is 0.69, and the CCC score is 0.60. With the GBDT machine learning residual correction, these values increase to 0.76 and 0.72, corresponding to improvements of 9.6$\%$ and 21.0$\%$, respectively. The increase in the Pearson coefficient indicates that the corrected reconstruction more closely follows the variations of the true velocity field, and the larger improvement in the CCC shows that the correction also suppresses systematic bias and better recovers the overall velocity amplitude.

We further quantify the reconstruction in Fourier space in Fig.~\ref{fig:box_pscorr}. The upper panel shows the momentum power spectrum $P_{qq}(k)$ for the true field, the linear reconstruction, and the GBDT-corrected reconstruction. We find that the linear reconstruction systematically underestimates the true power spectrum at all scales, and the discrepancy becomes increasingly larger at smaller scales. On the other hand, the machine learning corrected reconstruction follows the true power spectrum much more closely, indicating that the machine learning model is able to recover information beyond that captured by the linear approximation alone. The lower panel shows the correlation coefficient defined in Eq.~(\ref{eq:rqk}) as a function of $k$. On the largest scales, both reconstructions remain highly correlated with the true field, as expected. However, the linear reconstruction degrades steadily toward smaller scales, with the correlation dropping significantly in the nonlinear regimes. The GBDT corrected reconstruction maintains a higher correlation across the full $k$ range, with a slower decline at smaller scales. The result indicates that the machine learning correction preserves the large-scale information already present in the linear reconstruction while extending a better reconstruction to smaller scales.

\begin{table*}
\centering
\setlength{\tabcolsep}{12pt}
\renewcommand{\arraystretch}{1.15}
\begin{tabular}{|c|c|c|c|lcc|}
\hline
Direction & Tracer & Redshift &$\sigma_z/(1+z)$ & Method &  Correlation, $r_v$ & CCC, $\rho_v$ \\
\hline
\multirow{16}{*}{LOS}& \multirow{12}{*}{LRG} & \multirow{9}{*}{$0.4 < z < 0.6$ } & \multirow{5}{*}{0.00} & Linear  & 0.63 & 0.55 \\
&&&& GBDT  & 0.89 & 0.88 \\
&&&&  Transformer & 0.91 & 0.90 \\
&&&&  Transformer (Disp. only) & 0.73 & 0.69 \\
&&&&  Transformer (Core only) & 0.90 & 0.89 \\
\cline{4-7}
&&& \multirow{2}{*}{0.01}& Linear   & 0.41 & 0.30 \\
&&&& Transformer   & 0.56 & 0.47 \\
\cline{4-7}
&&& \multirow{2}{*}{0.02} &  Linear   & 0.27 & 0.24 \\
&&&&  Transformer   & 0.51 & 0.42 \\
\cline{3-7}
&& \multirow{3}{*}{$0.7 < z < 0.9$} & \multirow{3}{*}{0.00}&  Linear  & 0.65 & 0.55  \\
&&&& GBDT   & 0.90 & 0.89 \\
&&&&  Transformer  & 0.92 & 0.91 \\
\cline{2-7}
&\multirow{4}{*}{ELG} & \multirow{2}{*}{$0.7 < z < 0.9$} & \multirow{2}{*}{0.00} & Linear & 0.57 & 0.53  \\
&&&& Transformer  & 0.83 & 0.82 \\
\cline{3-7}
&&\multirow{2}{*}{$1.0 < z < 1.2$}& \multirow{2}{*}{0.00}  & Linear  & 0.37 & 0.36  \\
&&&& Transformer   & 0.76 & 0.74 \\ \hline
\multirow{6}{*}{ Transverse} & \multirow{6}{*}{LRG}& \multirow{6}{*}{$0.4 < z < 0.6$ } & \multirow{2}{*}{0.00} & Linear & 0.67 & 0.59 \\
&&&& Transformer & 0.85 & 0.84 \\
\cline{4-7}
&&& \multirow{2}{*}{0.01}& Linear  &  0.47 & 0.43 \\
&&&& Transformer  & 0.72 & 0.68 \\
\cline{4-7}
&&&   \multirow{2}{*}{0.02}& Linear  &  0.45 & 0.40 \\
&&&& Transformer  & 0.65 & 0.59 \\
\hline
\end{tabular}
\caption{A summary of the reconstruction performance in the lightcone setting, quantified by the Pearson correlation coefficient, $r_v$, and the concordance correlation coefficient (CCC), $\rho_v$. Results are shown for LRG samples in the redshift bins $0.4<z<0.6$ and $0.7<z<0.9$, and for ELG samples in the redshift bins $0.7<z<0.9$ and $1.0 < z < 1.2$. We compare the baseline linear reconstruction with GBDT and Transformer models. The table also includes results for LOS and transverse velocity components under conditions with perfect redshift information and with photometric survey-like redshift uncertainties of $\sigma_z/(1+z)=0.01$ and 0.02. All analyses are conducted using the full set of features, except for two lower redshift LRG sample which use subsets of the features: one uses just displacement features (Disp. only) and a second uses just simple scalar and vector displacement and environment features (Core only) as described in Sec.~\ref{sec:feature}. 
}
\label{tab:results}
\end{table*}

\begin{figure*}[!t]
    \centering
    \subfigure[LRG $v_{\parallel}$ $0.4 < z < 0.6$]{\includegraphics[width=\textwidth]{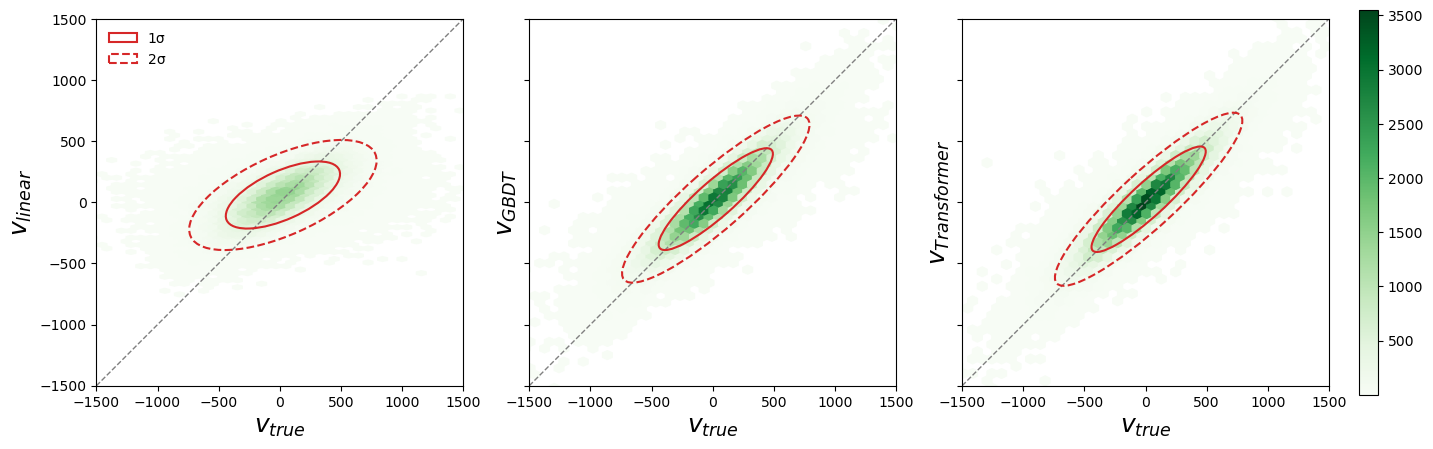}}\\
    \subfigure[LRG $v_{\parallel}$ $0.7 < z < 0.9$]{\includegraphics[width=\textwidth]{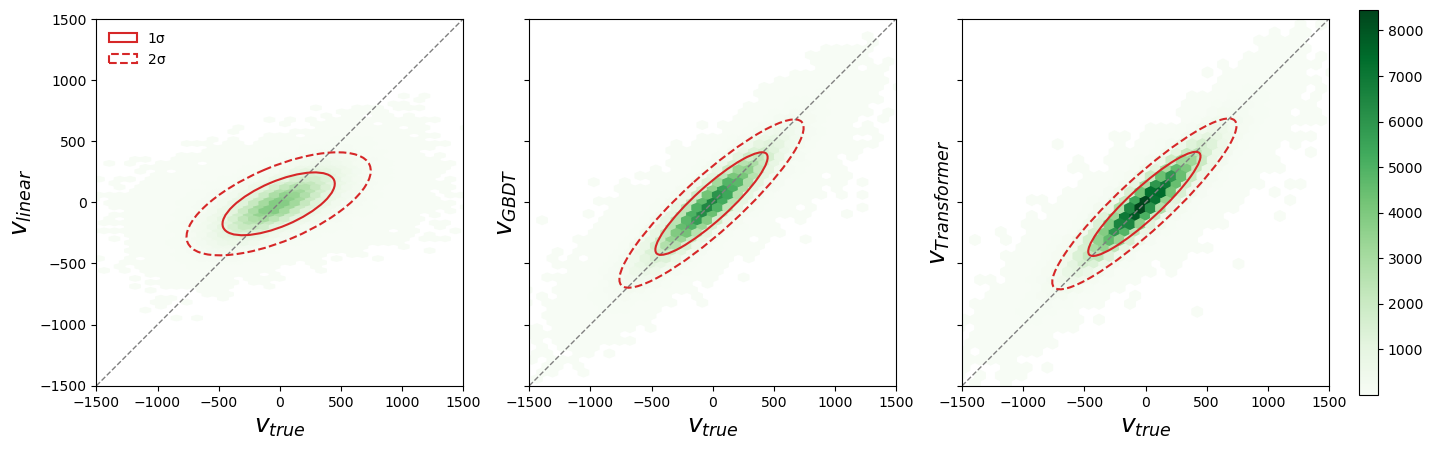}}
    \caption{Same as Fig. \ref{fig:box_velocity} but for lightcone samples centered at $z=0.5$ [top] and $z=0.8$ [bottom]. In each row, the three panels show the baseline linear reconstruction [left], the GBDT residual corrected reconstruction [middle], and the Transformer residual corrected reconstruction [right]. 
    }
    \label{fig:lightcone_velocity}
\end{figure*}

\begin{figure}[t!]
    \centering
    \includegraphics[width=0.48\textwidth]{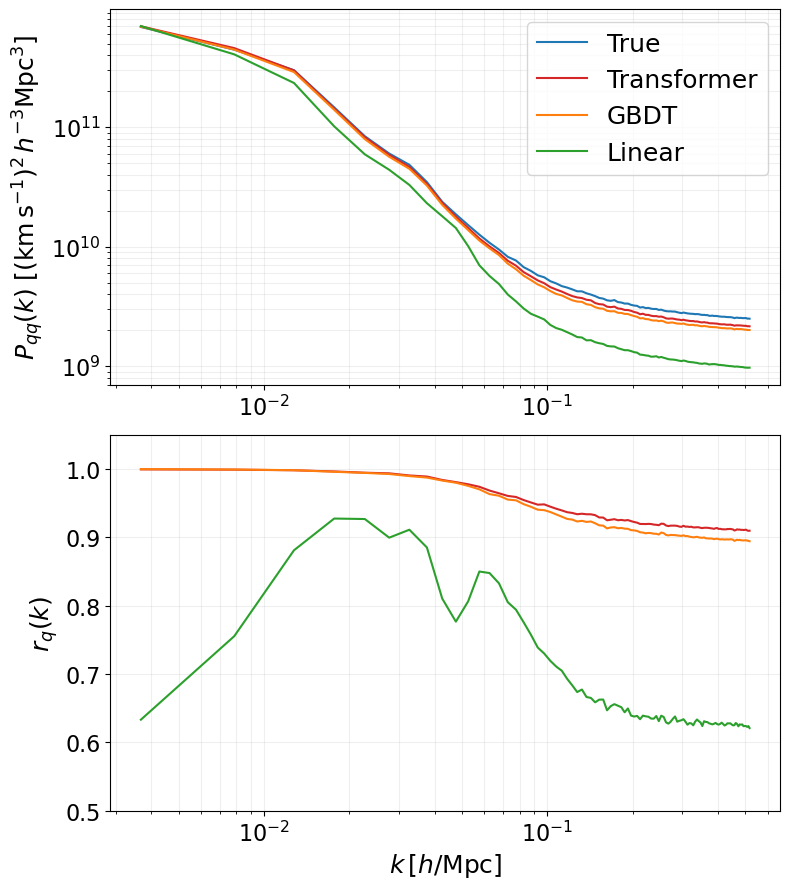}
    \caption{Same as Fig.~\ref{fig:box_pscorr}, but presents the comparison of the true [blue], Transformer-corrected [red], GBDT-corrected [orange], and linear reconstructed [green] momentum fields in the lightcone samples centered at $z=0.5$.
     }
    \label{fig:lightcone_pscorr}
\end{figure}

\subsection{Lightcone Reconstruction}
\label{sec:lightcone}

We now turn to the more realistic lightcone catalogs, where the reconstruction is performed in the presence of survey geometry, a local line-of-sight definition, and redshift evolution across the sample. We evaluate the performance in two DESI-like LRG redshift bins centered at $z=0.5$ and $z=0.8$. We compare the baseline linear reconstruction with two residual learning models, GBDT and Transformer. Fig.~\ref{fig:lightcone_velocity} shows the reconstructed LOS velocities against the true values for the two lightcone samples. In both redshift bins, the linear reconstruction captures the overall trend of the velocity field, but the scatter around the one-to-one relation remains substantial and the reconstructed amplitudes are visibly compressed. After applying either machine learning correction, the distribution becomes much more concentrated along the diagonal, with both reduced scatter and improved amplitude. The improvement is strongest for the Transformer, whose contours are systematically tighter and more closely aligned with the one-to-one relation than those of the GBDT model.

Quantitatively, the two lightcone redshift bins give highly comparable results as shown in Table. \ref{tab:results}. For the $0.4 < z < 0.6$ sample, the linear reconstruction yields a correlation coefficient of 0.63 and a CCC score of 0.55. The GBDT corrected reconstruction improves these to 0.89 and 0.88, while the Transformer further improves them to 0.91 and 0.90. For the $0.7 < z < 0.9$ sample, the corresponding values improve from 0.65 and 0.55 for the linear reconstruction to 0.90 and 0.89 for GBDT, and to 0.92 and 0.91 for the Transformer. Relative to the linear baseline, this corresponds to an improvement of $41\%$ in correlation and $60\%$ in CCC for GBDT , and $44\%$ and $64\%$ for the Transformer at $0.4 < z < 0.6$. At $0.7 < z < 0.9$, the GBDT improves $39\%$ and $62\%$, while the Transformer improves $42\%$ and $65\%$. The final performance is therefore comparable at the two redshifts, and in both bins the Transformer gives the best reconstruction.

Compared with the box reconstruction in Sec.~\ref{sec:box}, the lightcone linear reconstruction yields slightly lower correlation and CCC values. This is expected as survey geometry, local LOS effects, and redshift evolution are included. However, after applying the machine learning residual correction, the final lightcone performance surpasses the box, with both the correlation coefficient and the CCC reaching about 0.9 in the two redshift bins. We attribute this improvement to the fact that the ML task in the box setup is defined only for the velocity component along a single fixed axis ($z$), whereas the lightcone case effectively samples a much broader range of directional configurations because the local line of sight varies across the survey footprint. Since the underlying velocity field is intrinsically three dimensional, this richer directional information makes the residual correction more learnable and improves the final performance.

We then test the Fourier-space statistics and the results are shown in Fig.~\ref{fig:lightcone_pscorr}. In both redshift bins, the linear reconstruction underestimates the true momentum power spectrum at all scales, with the discrepancy increasing both toward larger and smaller scales. The performance on larger scales is purely a consequence of the lightcone being a shell over a limited redshift range while at smaller scales it is because of poor characterization of the nonlinear velocity. By contrast, both machine learning reconstructions track the true spectrum much more closely, substantially reducing the amplitude mismatch. This improvement does not imply that the machine learning models reconstruct genuinely new large-scale modes that are absent from the linear estimator. Rather, on these scales the linear reconstruction already captures most of the phase information in the velocity field, while the lightcone geometry and the finite radial width of the shell primarily introduce a scale-dependent response and amplitude mismatch in the reconstructed momentum field. Since the machine learning models are trained to predict the residual with respect to the linear reconstruction, they act as a calibration of this response rather than a replacement of the large scale reconstruction. The input features include the local LOS direction, redshift, and multiscale displacement and density derived quantities, allowing the models to learn systematic lightcone and geometry dependent residual patterns. As a result, the machine learning corrected fields can recover the large scale momentum power more accurately while preserving the large scale correlations already present in the linear reconstruction. The Transformer is generally the closest to the true power spectrum, with GBDT following closely behind, while the linear result remains systematically low across the full $k$ range. The behavior is very similar at $0.4 < z < 0.6$ and $0.7 < z < 0.9$, showing that the learned corrections can capture residual patterns that remain robust across different redshifts. From the lower panels of Fig.~\ref{fig:lightcone_pscorr}, we find that, on large scales, both machine learning reconstructions remain nearly perfectly correlated with the true momentum field, while the linear reconstruction is noticeably less correlated. Toward higher $k$ ($>\sim 0.02h$/Mpc), the linear result degrades rapidly, the GBDT prediction declines more slowly, and the Transformer retains the highest correlation over most of the plotted range ($\sim0.04<k<0.4h$/Mpc). At $k$ well into the nonlinear regime, $\sim 0.3-0.4\,h\,\mathrm{Mpc}^{-1}$, the Transformer achieves $r_q\simeq 0.91$ (0.92) for the lower (higher) LRG redshift bin, compared with about 0.90 (0.89) for GBDT and about 0.63 (0.66) for the linear reconstruction. This resilience in the reconstruction down to scales of a few Mpc closely mirrors what we found in the box case: machine learning residual correction preserves the large-scale information already present in the linear estimator while extending the reconstruction fidelity further into the nonlinear regime. 

We also test a simplified Transformer model that uses only the displacement features. In this setting, the input contains only the core features from the reconstructed displacement features at the three smoothing scales, while all environment-related features discussed in Sec.~\ref{sec:feature} are excluded. The main motivation for this test is robustness in realistic applications. Many of the environment related quantities are constructed from spatial derivatives of the reconstructed fields. While these derivative features can be informative in simulations, we are concerned that they may become noisier in real observations. This is particularly relevant in lightcone settings, where survey geometry, sparse sampling, and observational systematics can amplify noise in derivative based quantities. It is therefore useful to examine how well the Transformer performs when restricted to only the most basic and potentially more stable reconstruction features. For the LRG sample with redshift $0.4 < z < 0.6$, the Transformer model can still achieve a correlation coefficient of $0.73$ and a CCC score of $0.69$. These correspond to relative improvements of $15.6\%$ and $27.5\%$, respectively, over the linear reconstruction baseline. This indicates that the Transformer gain is not solely driven by the additional environment descriptors. Instead, the basic multiscale displacement features already contain substantial information that can be effectively used by the Transformer. 

We further test a second reduced feature set that keeps only the simplest displacement and environment (density-derived) information. This set includes the linear displacement features, the displacement magnitude, the LOS displacement projection and its magnitude, together with the scalar density, the density gradient, and the density gradient magnitude at the three smoothing scales. The corresponding cross scale difference features are also included. Compared to the full feature set, this test excludes the nonlinear displacement features and the more complex tensor based environment features, such as the Hessian and tidal tensor quantities. It therefore provides an intermediate case between the displacement-only model and the full model, allowing us to assess whether the simpler density gradient information is robust and informative without relying on the potentially noisier higher order tensor features. For the LRG sample with redshift $0.4 < z < 0.6$, this reduced model achieves a correlation coefficient of $0.90$ and a CCC score of $0.89$ respectively. This performance is very close to that of the full feature set, indicating that much of the Transformer's information gain comes from the combination of multiscale displacement information with the simplest density-derived features.

Beyond the LRG samples discussed above, we also test the reconstruction performance for the ELG tracers in the lightcone setting. The results are shown in Table.~\ref{tab:results} for the redshift bins $0.7<z<0.9$ and $1.0<z<1.2$. For the ELG sample at $0.7<z<0.9$, the Transformer substantially improves over the linear reconstruction, as it did for the LRGs in the same redshift range, increasing the Pearson correlation coefficient from $0.57$ to $0.83$ and the CCC score from $0.53$ to $0.82$. This demonstrates that the machine learning model can also recover a significant amount of nonlinear velocity information for the ELGs even while they are a fundamentally different tracer of the cosmic density. Compared with the LRG sample in the same redshift bin, the LRG reconstruction is slightly more accurate than the ELG reconstruction. The lower ELG performance is caused by the lower tracer bias, number density, and sampling properties of the ELG population \cite{Hadzhiyska:2023nig}, but the relative gain over the linear baseline remains large. 

In the higher redshift bin, $1.0<z<1.2$, the ELG linear reconstruction becomes  less accurate, with $(r_v,\rho_v)=(0.37,0.36)$. Nevertheless, the Transformer model still recovers a significantly larger fraction of the missing velocity information, improving the performance to $(r_v,\rho_v)=(0.76,0.74)$. This corresponds to an even larger relative gain than in the lower redshift ELG bin, although the final reconstruction accuracy remains lower. 

The relatively lower performance of linear reconstruction, for ELGs relative to LRGs in the same redshift bin, and for the ELGs in the higher vs lower redshift sample, is expected, since its effectiveness is closely related to the combination $\bar{b}^2 n(z)$, where $\bar{b}$ is the mean galaxy bias across the redshift bin and $n(z)$ is the number density. For the LRG sample, based on the values in Table~\ref{tab:galnum}, $b(z)^2 n(z) \approx$ 3.0 and 3.4$\times 10^{-3}$ $h^3$/Mpc$^3$ in the redshift ranges $0.4<z<0.6$ and $0.7<z<0.9$, respectively. In contrast, the ELG samples at $0.7<z<0.9$ and $1.0<z<1.2$, respectively have $b(z)^2 n(z)$ = 0.49 and 0.41 $\times 10^{-3}$ ($h$/Mpc)$^3$. These differences explain the relative performance of linear reconstruction discussed above. The LRG samples, with substantially larger $\bar{b}(z)^2 n(z)$ values, provide a better estimate of the displacement field and therefore benefit more from the reconstruction procedure. The lower effective tracer densities of the ELG samples, on the other hand, lead to the reconstructed displacement field being noisier, reducing the improvement that can be achieved. 

Our findings demonstrate that the Transformer's nonlinear residual corrections can be successfully implemented to a variety of tracers while recognizing that the performance is sensitive to the tracer number density and bias both of which can be redshift dependent.

In addition to the LOS velocity reconstruction, we also consider how the Transformer can be applied to accurately estimate the perpendicular component. We train a separate Transformer model using features constructed specifically for that direction. The model uses the same feature set as in the LOS case, but all direction dependent quantities are defined with respect to the perpendicular component rather than the LOS component. The corresponding results without redshift errors are shown in Table. \ref{tab:results} for the LRG sample at $0.4<z<0.6$. The linear reconstruction achieves $(r_v,\rho_v)=(0.67,0.59)$. The Transformer correction improves this substantially to $(r_v,\rho_v)=(0.85,0.84)$. This demonstrates that the residual correction is not limited to the LOS velocity, but can also recover a significant fraction of the missing transverse velocity information. Compared with the LOS result in the same redshift bin, the Transformer reconstructed perpendicular velocity has a slightly lower Pearson correlation, $r_v=0.85$ compared to $0.91$ for the LOS component. Nevertheless, the performance remains high, and the gain over the linear baseline is comparable in size. These results support the use of Transformer based velocity reconstruction for applications that require perpendicular velocities such as the moving lens analyses.
\subsection{Redshift Uncertainties}
\label{sec:Zerr}

We examine how redshift uncertainties impact our velocity reconstruction, using the $0.4 < z < 0.6$ LRG sample as the reference. We model redshift uncertainties by adding Gaussian noise to the true redshifts, and adopt $\sigma_z/(1+z)$ = 0.01 and 0.02, which respectively reflect the stretch and target goals for LSST photo-z errors \cite{lsstsciencecollaboration2009lsstsciencebookversion}. This also helps us to bound the effect of redshift uncertainties from slitless spectroscopy surveys from Euclid and Roman, expected to be $\sigma_z/(1+z) \sim$ 0.001 \cite{Wang:2010gq}. 

The reconstruction results for the LOS and perpendicular components of the peculiar velocity are summarized in Table~\ref{tab:results}. We find that redshift uncertainties degrade the reconstruction performance in both the LOS and perpendicular directions, with a stronger impact on the LOS component. Similar results are reported in \cite{Hadzhiyska:2023nig} for the linear reconstruction. For $\sigma_z/(1+z)=0.01$, the LOS correlation coefficient decreases to $r=0.56$, while the perpendicular component remains comparatively better reconstructed, with $r=0.72$. When the redshift uncertainty is increased to $\sigma_z/(1+z)=0.02$, the degradation becomes more pronounced, with $r$ further decreasing to 0.51 for the LOS component and 0.65 for the perpendicular component. A similar trend is observed for the CCC, which decreases from 0.47 to 0.42 for the LOS direction and from 0.68 to 0.59 for the perpendicular direction.

This stronger degradation for reconstruction along the LOS is to be expected since the redshift errors smear the galaxy distribution, erasing small scale structures and weakening the inferred density gradients, used in reconstruction along that axis. In contrast, the transverse information is less directly affected by redshift inaccuracies, making its reconstruction more robust. The degradation from redshift uncertainty affects both stages of the reconstruction pipeline both the linear reconstruction and the nonlinear machine learning residual estimation. The ML model is trained on features derived from the degraded reconstructed linear field and cannot fully recover information that has been erased by photometric redshift error. Nevertheless, the Transformer correction remains beneficial relative to the corresponding linear baseline.

While photometric redshift uncertainties do lead to a loss in velocity reconstruction accuracy relative  to the idealized spectroscopic case, it is conceivable that the higher galaxy number density could ameliorate the degradation through enabling more precise gradient estimation.  DESI's multi-object fiber spectroscopy \cite{DESI:2016fyo} provides highly accurate redshift measurements but their sample sizes/number densities are typically much smaller than those available from photometry, such as from the Rubin Observatory LSST. The DESI Main LRG galaxy sample has $n(z)\approx 0.17$ galaxies per sq. arcmin. while LSST is expected to have observe 20-40 galaxies per sq. arcmin., roughly two orders of magnitude greater.  The signal-to-noise ratio (SNR) of kSZ analysis might roughly scale as $\sim r_v\sqrt{N}$, for a sample size, $N$ and the velocity reconstruction correlation coefficient, $r_v$. These results indicate that machine learning reconstruction of both LOS and transverse velocity components could be viable in the presence of redshift uncertainties as the reduction in $r_v$  could be countered by a commensurate increase in sample size.

\section{kinematic SZ Applications}
\label{sec:kSZ_results}
In this section, we discuss results of applications of the reconstructed velocities to two common kSZ analyses. We assess the recovery of the pairwise velocity statistic in Section. \ref{sec:pairwise} and examine the stacked kSZ profile in Section \ref{sec:kSZ_profile_results}.

\subsection{Pairwise Velocity Reconstruction}
\label{sec:pairwise}

We apply the reconstructed velocities to determine the pairwise velocity statistic described in Section~\ref{sec:pairwise} which probes the relative infall pattern of galaxy pairs across a wide range of comoving separations. It provides a useful study of how errors in the reconstruction might propagate into a practical large-scale structure observable.

kSZ analyses do not require perfect recovery of individual galaxy velocity, but rather an accurate reconstruction of the pairwise statistic averaged over the galaxy sample within the spatial separate bin. 

\begin{figure}
    \centering
    \includegraphics[width=\columnwidth]{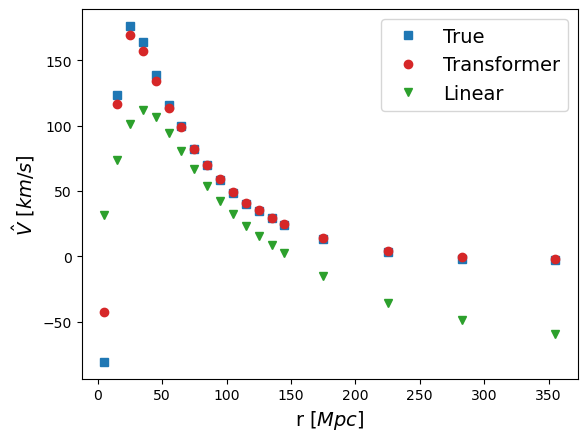}
    \caption{Comparison of the true [blue],  linear-reconstructed [green] and Transformer-corrected [red] pairwise velocity statistic, $\hat{V}$, as a function of comoving separation, $r$.
    }
    \label{fig:pairwise}
\end{figure}

The results are shown in Fig.~\ref{fig:pairwise}, where we compare the pairwise signal measured from the true halo peculiar velocities, the linear reconstruction, and the Transformer-corrected reconstruction for the $0.4 < z < 0.6$ LRG sample. The linear reconstruction is able to capture the overall scale dependence of the pairwise statistic, but it systematically underestimates the amplitude over the full separation range by roughly 45\%. By contrast, the Transformer model can recover the true pairwise velocity much more closely. Over most of the range, the Transformer successfully recovers the nonlinear information that is relevant for pairwise motions on scales with $r > 45\,\mathrm{Mpc}$ to within $2.5\%$ of the true signal.  
Our results show that the machine learning correction improves not only the object-level velocity prediction but also the pairwise signal.

We consider a specific way in which these pairwise velocities can be used, to perform an optical depth reconstruction test with kSZ measurements, based on Eq.~(\ref{eq:ksz_halo}).
To estimate the kSZ temperatures, $\Delta T_{kSZ,i}(\theta)$, and to mirror real observational analysess, we perform kSZ aperture photometry measurements on a simulated kSZ map, centered on the LRG sample, assumed to serve as proxies to the centrs of galaxies clusters and groups.

Using the pairwise velocity estimates for the linear and Transformer-derived velocities, we find best-fit mass-averaged optical depths, $10^4\bar{\tau}$ = $0.34 \pm 0.07$ and
$0.67 \pm 0.09$, respectively, in comparison to that obtained from the true pairwise velocity of $0.66 \pm 0.08$. 
 The true and machine learning velocity estimates give consistent optical depths, while the linear reconstructed velocity leads to a significantly lower and biased inferred optical depth.

\subsection{Stacked kSZ Profile}
\label{sec:kSZ_profile_results}
\begin{figure}
    \centering
    \includegraphics[width=\columnwidth]{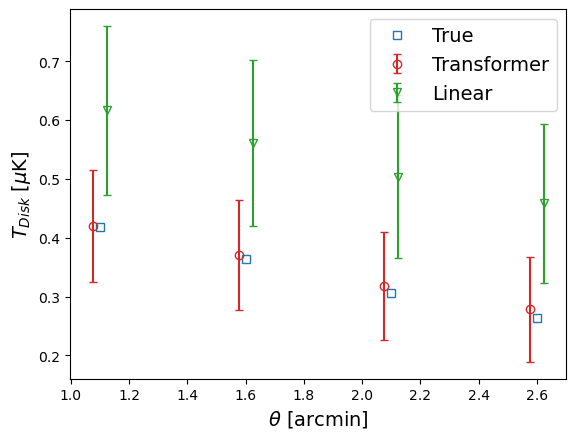}
    \includegraphics[width=\columnwidth]{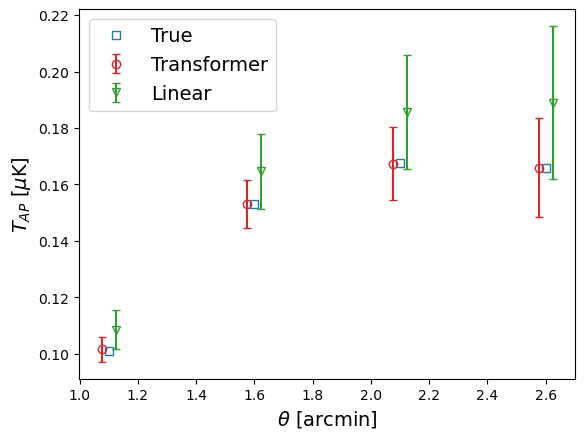}
    \caption{Stacked kSZ temperature profiles as a function of aperture radius $\theta$. We compare the profile constructed from the true halo velocities [blue], the Transformer-corrected reconstruction [red], and the linear reconstruction [green] for the kSZ signal using the disk kSZ temperature [top] and aperture photometry [bottom]. To reflect the real observation analysis, we add realistic primary CMB and instrument noise on top of the kSZ signal. 1$\sigma$ uncertainty estimated from the bootstrap analysis is also included.}
    \label{fig:ksz_profile}
\end{figure}

We examine the application of the reconstructed velocities to recover the stacked kSZ profile as defined in Eq.~(\ref{eq:kSZ_profile}) for the mock LRG sample in $0.4 < z < 0.6$. The correlation coefficient is estimated from the simulations, as is also done in the analysis of real data \citep{Hadzhiyska:2026uyq}. The correlation function is found to be robust, changing only by 0.5\% (from $r_{corr}=0.626$ to 0.623 when calculated using only 30\% of the data rather than the full sample).  In forecasting by the Simons collaboration for joint analyses with the full DESI data (\citep{SimonsObservatory:2018koc}), they assumed 2.9M LRGs measured in an overlapping area of 10,000 sq. deg. Our analysis here, focused on $\sim~750k$ galaxies, is therefore a conservative one. 

The results are shown in Fig.~\ref{fig:ksz_profile}.  We compare stacked kSZ profiles constructed using the true halo peculiar velocities, the linear reconstruction, and the Transformer-based reconstruction.  We consider two commonly used methods to measure the kSZ signal (e.g. \citep{Gong:2023hse}), the aperture photometry $T_{AP}$ and the mean temperature inside a disk, $T_{Disk}$. We calculated the profile at four apertures: 1.1$^\prime$, 1.6$^\prime$, 2.1$^\prime$, and 2.6$^\prime$. We find that the linear reconstruction result systematically overestimates the amplitude at all apertures. The Transformer-corrected reconstruction accurately recovers the true stacked kSZ profiles. The agreement is more pronounced at the intermediate apertures for the aperture photometry measurements, where the Transformer differs from the true values by less than 1$\%$. The error bars shown in Fig.~\ref{fig:ksz_profile} reflect the uncertainties in the simulated kSZ signal, due to residual primary CMB and instrument noise; the uncertainties in velocity reconstruction are negligible in comparison.

These results demonstrate that the machine learning correction improves not only the reconstruction of individual halo velocities and pairwise velocity statistic, but also the stacked kSZ profiles. Overall, the Transformer-based residual correction is able to preserve the physically relevant information required for kSZ profile measurements while significantly reducing the systematic bias present in the linear reconstruction.
\section{Conclusion}
\label{sec:conclusion}

In this work, we show how machine learning can improve peculiar velocity reconstruction beyond the standard linear framework for DESI- and Rubin-like LRG and ELG samples. Starting from a physically motivated linear reconstruction baseline, we train two residual learning models, a GBDT and a Transformer, to predict corrections to the reconstructed LOS velocity. We build on the standard linear reconstruction pipeline by learning the structured nonlinear residuals. 

We demonstrate that the machine learning reconstruction increases both the Pearson correlation coefficient ($r_v$) and the concordance correlation coefficient (CCC, $\rho_v$), the latter provides a clear measure of how well the overall amplitude of the reconstructed velocity agrees with the true value. The machine learning predictions also more accurately recover the momentum power spectrum and cross-correlation coefficient at all sampled scales. These results show that the machine learning model can effectively recover information beyond the linear approximation while preserving the large-scale physical information already captured by the linear baseline reconstruction.

For the periodic box snapshot at $z=0.5$, the GBDT model provides a clear improvement over the linear baseline in both real and Fourier space metrics. For the LRG sample the linear LOS velocity estimate has $(r_v,\rho_v)=(0.69,0.60)$ which increases to (0.76,0.72) with the GBDT residuals. 

The improvements are more significant in the realistic lightcone setting. For the $0.4<z<0.6$ LRG sample, the linear reconstruction gives $(r_v,\rho_v)=(0.63,0.55)$, which improves to $(0.89,0.88)$ for GBDT and $(0.91,0.90)$ for the Transformer. The improvement in the CCC, $\rho_v$,indicates  not only  a tighter linear relationship between the reconstructed and true velocities but also that the machine learning predictions more accurately match the full amplitude of the velocity, while the linear estimate systematically underestimates it. For the $0.7<z<0.9$ LRG sample, the corresponding values improve from $(0.65,0.55)$ to $(0.90,0.89)$ and $(0.92,0.91)$, respectively. In both redshift bins, the machine learning reconstructions can also recover the LOS momentum power spectrum much better than the linear baseline and maintain substantially higher cross-correlation with the true field in the nonlinear regime. Among the two models used here, the Transformer gives the best overall performance, indicating that a more complicated architecture is able to extract additional information from the multi-scale reconstruction features. 

We also find similar results for the ELG sample in the $0.7 < z < 0.9$ redshift bin. The transformer model also improves the corresponding values from (0.57,0.53) to (0.83,0.82) respectively. For the higher-redshift ELG bin $1.0<z<1.2$, the linear model gives a very poor approximation to the  true velocity, with $(r_v,\rho_v)=(0.37,0.36)$. The Transformer model, however, improves the reconstruction significantly with $(r_v,\rho_v)=(0.76,0.74)$. These results show that although the reconstruction performance degrades at higher redshift, the learned residual corrections remain effective for ELG tracers. We discuss how the expected performance can be heuristically related to bias and number density of the galaxy samples, $b^2(z)n(z)$.
 
We also tested a simplified Transformer model that uses only the most basic reconstruction features, excluding the derivative based environment features. Even in this reduced setting, the model can still improve significantly over the linear baseline. This is encouraging for realistic survey applications, where derivative based features may be noisier and less robust than in simulations.

We further quantified the impact of redshift uncertainties in redshift identification degrade the reconstruction. A stronger effect is seen on the LOS component than on the transverse one relevant to the moving lens analysis. Our results highlight the trade off between galaxy sample size and redshift accuracy in velocity reconstruction performance and offer the possibility that the higher number density of photometric galaxy samples may counter the reduction in the reconstructed velocity correlation. 

We also examined the impact of the improved reconstruction on downstream kSZ observables. For the pairwise velocity statistic, we find that the linear reconstruction can recover the overall scale dependence but systematically underestimate the true amplitude, whereas the Transformer corrected results can recover the true pairwise signal accurately. In particular, on linear scales with $r>45$ Mpc, the average difference from the true pairwise velocity is below $2.5\%$. We also find a similar trend for the stacked kSZ profile measured from the simulated kSZ map using both aperture photometry and disk kSZ temperatures. The linear reconstruction systematically overestimates the profile amplitude, while the Transformer corrected result recovers the true profile accurately, with differences below $1\%$ at intermediate apertures. These results show that the machine learning correction improves not only object level velocity reconstruction, but also improves the kSZ observables that are directly relevant for cosmological analysis.

The machine learning model used in this work is trained assuming a fixed cosmology used to produce the training simulations. 
The learned nonlinear residuals could in principle depend on the cosmological model. Quantifying the sensitivity of the reconstruction performance to changes in cosmological parameters would be valuable for assessing how such methods can be used not only to improve velocity reconstruction, but also to test cosmological models. We leave a systematic test of both the cosmology dependence and a fuller assessment of the potential to use this approach for photometric redshift surveys to future work.

In summary, this work demonstrates that the machine learning models can provide an effective and practical way to improve peculiar velocity reconstruction beyond the linear approximation. The method is shown to be promising in realistic lightcone settings relevant for ongoing and upcoming LSS and CMB surveys. 
 We focused here on a minimal set of displacement and number density-based features using DESI-like spectroscopic galaxy samples. The advent of LSST, Roman and Euclid LSS survey data and SO and FYST CMB maps provides rich opportunities to consider a far broader set of input features, larger galaxy samples and a broad array of applications. Valuable avenues for future study include assessing improvements in  velocity-based constraints on the growth rate of structure and modified gravity, the measurement of large scale cosmic flows, and velocity cross-correlations with CMB lensing, galaxy lensing, cluster catalogs, and intensity mapping surveys. 

\begin{acknowledgments}
We thank Boryana Hadzhiyska for helpful conversations during the course of this research and thank Patricio Gallardo for making publicly available the pairwise correlation code used in this work. We also acknowledge the use of the National Energy Research Scientific Computing Center (NERSC) computational resources and platform, through an allocation provided to the DESI collaboration, in this work. The work of YG and RB is supported by NSF grant AST-2206088, NASA grant 22-ROMAN11-0011 and NASA grant 12-EUCLID12-0004.

\end{acknowledgments}
\appendix
\section{Simulated kSZ map creation and analysis}
\label{app:kSZ}
\subsection{Simulated kSZ Map}
\label{subsec:kSZ_map}
To assess how velocity reconstruction errors propagate into a practical kSZ analysis, we construct a mock kSZ temperature map from the halo lightcone catalog, following \citep{Roper:2025ygs}. In the non-relativistic limit, the kSZ temperature fluctuation along the line of sight $\hat{\mathbf n}$ can be written as
\begin{equation}
\frac{\Delta T_{kSZ}(\hat{n})}{T_{0}}
=
-\frac{\sigma_{T}}{c}
\int n_e\, v_{los} \, {d}l,
\label{eq:ksz}
\end{equation}
where $\sigma_{T}$ is the Thomson cross section, $n_e$ is the free electron number density, and $v_{los}$ is the LOS peculiar velocity.

We model the kSZ contribution from each halo $i$ as a projected optical depth profile multiplied by its bulk LOS velocity,
\begin{equation}
\Delta T_i(\theta)
=
- T_{\mathrm{CMB}}\,
\tau_i(\theta)\,
\frac{v_{\parallel,i}}{c},
\label{eq:ksz_halo_app}
\end{equation}
where $\theta$ is the angular distance from the halo center. The optical depth is written as
\begin{equation}
\tau_i(\theta)
=
\frac{N_{e,i}\sigma_T}{D_{A,i}^2}\,
W_G(\theta;\Sigma_i),
\label{eq:tau_profile}
\end{equation}
where $N_{e,i}$ is the total number of free electrons, $D_{A,i}$ the angular diameter distance, and
\begin{equation}
W_G(\theta;\Sigma_i)
=
\frac{1}{2\pi \Sigma_i^2}
\exp\left(
-\frac{\theta^2}{2\Sigma_i^2}
\right),
\qquad
\int W_G  \, d\Omega= 1.
\label{eq:gaussian_profile}
\end{equation}
The effective angular width is written as
\begin{equation}
\Sigma_i^2 = \sigma_B^2 + \sigma_{R,i}^2,
\label{eq:sigma_tot}
\end{equation}
where $\sigma_B$ is the instrumental beam size and $\sigma_{R,i}$ is the intrinsic angular extent of the gas profile. In our implementation, we use
\begin{equation}
\sigma_B
=
\frac{\theta_{\mathrm{FWHM}}}{\sqrt{8\ln 2}},
\qquad
\theta_{\mathrm{FWHM}}=2.1^\prime,
\label{eq:beam_sigma}
\end{equation}
and
\begin{equation}
\sigma_{R,i}
=
\alpha \frac{R_i}{D_{A,i}},
\label{eq:sigma_R}
\end{equation}
with $\alpha=0.6$. The characteristic halo radius is defined using a spherical overdensity of $\Delta=180$ with respect to the mean matter density,
\begin{equation}
R_i
=
\left[
\frac{3 M_i}
{4\pi \Delta \bar{\rho}_m(z_i)}
\right]^{1/3},
\qquad
\Delta=180.
\label{eq:R180b}
\end{equation}

Assuming fully ionized gas, the total number of free electrons can be calculated as
\begin{equation}
N_{e,i}
=
\frac{f_{gas} M_i}{\mu_e m_p},
\label{eq:Ne_def}
\end{equation}
where $m_p$ is the proton mass, $\mu_e=1.17$ is the mean molecular weight of free electron, and we adopt $f_{gas}=0.1574$.

The full kSZ map is then obtained by summing the contribution from all halos,
\begin{equation}
\Delta T_{kSZ}(\hat{n})
=
\sum_i \Delta T_i(\theta_i).
\label{eq:ksz_sum}
\end{equation}
We construct the map in Healpix format \cite{Gorski:2004by} with nside=8192. For computation efficiency, each halo only contributes within a radius of $5\Sigma_i$ around the halo center. We use the true halo LOS velocity $v_{true}$ when generating the map, so this simulated map represents the true kSZ signal without any velocity reconstruction error introduced.

It is important to note that the map constructed here is a semi-analytic simulation rather than a high fidelity kSZ simulation. The gas distribution around each halo is modeled with an idealized Gaussian profile, so the resulting map is not intended to reproduce all details of the true kSZ sky. It captures the dominant dependence of the signal on halo optical depth, angular scale, and LOS velocity, and is therefore sufficient for testing observables relevant to velocity reconstruction, including stacked kSZ measurements.

We also simulate primary CMB and instrument noise to test the model performance in anticipation of realistic CMB observing conditions. The primary CMB is simulated using CAMB \citep{2011ascl.soft02026L} using the same cosmology as the AbacusSummit simulation. We model the instrument noise as white noise with an amplitude of 6.3 $\mu K$-arcmin and apply a Gaussian beam with FWHM = 1.4$^\prime$ for Simons Observatory observation at 150 GHz \cite{SimonsObservatory:2018koc}.

\subsection{Pairwise kSZ Momentum and Optical Depth Fitting}
\label{subsec:AP}

The kSZ pairwise momentum is given by
 \begin{equation}\label{eq:pkSZ}
     P_{kSZ} (r,z_i)= -\frac{\sum_{ij}(\Delta \bar{T}_{AP,i} - \Delta \bar{T}_{AP,j})c_{ij}}{\sum_{ij}c_{ij}^2},
\end{equation}
where the sum is over all pairs of tracers with separations in a bin centered on $r$, and in a redshift bin centered on $z$. $\Delta \bar{T}_{AP,i}$ is the kSZ temperature obtained from simulated maps by using aperture photometry centered on the $i^{th}$ tracer. 

The aperture photometry temperature is defined as
\begin{equation}
\Delta\bar{T}_{AP}(\hat{n})
= \Delta\bar{T}_{{Disk}}(\hat{n}) - \Delta\bar{T}_{Ann}(\hat{n}),
\label{eq:TAP}
\end{equation}
where the disk averaged temperature within an aperture of radius $\theta$ is given by
\begin{equation}
\bar{T}_{{Disk}}(\theta)
=
\frac{1}{N_{disk}}
\sum_{i\in {disk}}
\Delta T_{kSZ,i},
\label{eq:Tdisk}
\end{equation}
where the sum runs over all $N_{disk}$ pixels inside a circular aperture of radius $\theta$ centered around the tracer location $\hat{n}$. Similarly, the average temperature within a typically adjacent annulus of equal area can be written as
\begin{equation}
\bar{T}_{Ann}(\hat{n})
= \frac{1}{N_{ann}}
\sum_{j\in ann(\theta,\sqrt{2}\theta)}
\Delta T_{kSZ,j},
\label{eq:TAnn}
\end{equation}
where $ann(\theta,\sqrt{2}\theta)$ denotes the annulus with inner radius $\theta$ and outer radius $\sqrt{2}\theta$ centered on $\hat{n}$, and $N_{ann}$ is the number of pixels in that annulus. 


The pairwise kSZ profile is related to the pairwise velocity, $V$, in Eq.~(\ref{eq:pairwise_estimator}, through
\begin{equation}
    P(r)
    =
    \frac{T_{0}}{c}\,
    \bar{\tau}\,
    \hat{V}(r),
    \label{eq:Pmod}
\end{equation}
where $\bar{\tau}$ denotes the effective mass-averaged optical depth of the sample. 
We estimate $\bar{\tau}$ by fitting the measured pairwise kSZ profile with a model constructed from a given pairwise velocity estimate. 
The best-fit value is obtained by minimizing
\begin{equation}
    \chi^2(\bar{\tau})
    =
    \sum_{i,j}
    \Delta P_i(\bar{\tau})
    C^{-1}_{ij}
    \Delta P_j(\bar{\tau}) ,
\end{equation}
where $C_{ij}$ is the covariance matrix of the pairwise kSZ measurement 
estimated from the bootstrap analysis and,
\begin{equation}
    \Delta P_i(\bar{\tau})
    =
    P_{obs}(r_i)
    -
    P_{model}(r_i;\bar{\tau}) .
\end{equation}
Here $P_{obs}$ is the measured pairwise kSZ profile, while $P_{model}$ is the prediction obtained using the chosen pairwise velocity correlation obtained by reconstruction in Eq.(\ref{eq:Pmod}) and with the trial value of $\bar\tau$. We apply this fitting procedure to three different velocity inputs: the true, linear reconstructed, and machine learning derived pairwise velocities, so that the resulting best-fit optical depths directly quantify how velocity reconstruction affects the inferred optical depth. Following \cite{Calafut:2021wkx}, we use scales $r > 20$ Mpc for our fitting process.

\clearpage
\bibliography{draft}

\end{document}